\documentclass[final,5p,times,twocolumn]{elsarticle}
\usepackage[hidelinks]{hyperref}
\usepackage{xcolor}

\usepackage[normalem]{ulem}
\usepackage{graphicx}
\usepackage{amsmath}   
\usepackage[amssymb]{SIunits}

\makeatletter
\let\ps@pprintTitle\ps@empty
\makeatother

\begin{document}

\begin{frontmatter}

\title{Development and characterization of hybrid MCP-PMT with embedded Timepix4 ASIC used as pixelated anode}

\author[1,2]{R. Bolzonella\corref{cor1}}\ead{riccardo.bolzonella@cern.ch}
\author[3]{J.~Alozy}
\author[3]{R.~Ballabriga}
\author[1]{N.~V.~Biesuz}
\author[3]{M.~Campbell}
\author[1,2]{V.~Cavallini}
\author[1]{A.~Cotta~Ramusino}
\author[1,2]{M.~Fiorini}
\author[1]{E.~Franzoso}
\author[1,2]{M.~Guarise}
\author[3]{X.~Llopart Cudie}
\author[1,2]{G.~Romolini}
\author[1]{A.~Saputi}
\cortext[cor1]{Corresponding author}

\affiliation[1]{organization={INFN, Division of Ferrara}, 
                 addressline={via G. Saragat 1},
                 postcode={44122}, 
                 city={Ferrara}, 
                 country={Italy}}
\affiliation[2]{organization={Dept. of Physics and Earth Science, University of Ferrara}, 
                 addressline={via G. Saragat 1},
                 postcode={44122}, 
                 city={Ferrara}, 
                 country={Italy}}
\affiliation[3]{organization={EP Department, CERN}, 
                 addressline={1211 Geneva 23},
                 country={Switzerland}}

\begin{abstract}
We present a novel single-photon detector based on a vacuum tube incorporating a photocathode, a microchannel plate (MCP), and a Timepix4 CMOS ASIC functioning as a pixelated anode. Designed to handle photon rates up to 1 billion per second across a $7\,\centi\meter^2$ active area, the detector achieves outstanding spatial and temporal resolutions of $5-10\,\micro\meter$ and below $50\,\pico\second$ r.m.s., respectively. 

The Timepix4 ASIC comprises approximately 230,000 pixels, each integrating analog and digital front-end electronics. This enables data-driven acquisition and supports data transmission rates up to 160 Gb/s. External FPGA-based electronics manage both configuration and readout.

In order to test the timing performance of the Timepix4 ASIC we performed preliminary characterization of an assembly bonded to a $100\,\micro\meter$ thick n-on-p silicon sensor using a pulsed infrared laser, which demonstrated a per-pixel timing resolution of $110\,\pico\second$, with cluster-based averaging methods improving to below $50\,\pico\second$.

Six prototype detectors incorporating different MCP stack configurations and end-spoiling depths were produced by Hamamatsu Photonics. We report on their characterization, including dark count rates, gain, and spatial and timing resolutions, assessed both in laboratory conditions and during a test-beam campaign at CERN's SPS facility.
\end{abstract}

\begin{keyword}
Photon-counting detector, Hybrid pixel detector, Timepix4 
\end{keyword}

\end{frontmatter}
\section{Introduction\label{sec1}}

4DPHOTON is a project funded by the European Research Council (ERC) under the European Union's Horizon 2020 research and innovation programme (Grant Agreement No. 819627). It is hosted by the Italian National Institute for Nuclear Physics (INFN) and coordinated by Principal Investigator Prof. Massimiliano Fiorini. The European Organization for Nuclear Research (CERN) and the University of Ferrara (UNIFE) also play key roles in the project.

The project's objective is to develop a novel large-area photodetector capable of detecting single photons with outstanding spatial and temporal resolution, while maintaining low noise levels even at room temperature \cite{bib:Fiorini_2018, Alozy:2021kqn}. These unprecedented performance characteristics address a major limitation in many photon imaging applications, which often lack detectors that can simultaneously provide high spatial and timing precision.

In addition to its compact form factor and high-rate capabilities, the detector holds great potential for high-energy physics applications. It offers an innovative solution for Ring Imaging Cherenkov detectors, enabling efficient particle identification in high-luminosity environments.

\section{Hybrid detector structure and operating principle description}

\begin{figure}[t]
\centering
\includegraphics[width=0.4\textwidth]{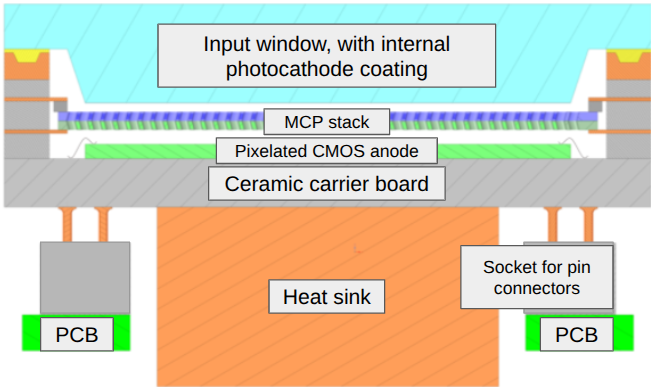}
\caption{Simplified cross-section of the hybrid MCP-PMT assembly.}
\label{fig:assembly}
\end{figure}

The "hybrid" assembly, whose cross-section is illustrated in Figure~\ref{fig:assembly}, is based on a vacuum tube that houses the components of the MCP-PMT under ultra-high vacuum conditions (approximately $10^{-10}\,\milli$bar). 

At the top of the tube, a silicate window—transparent to infrared, visible, and ultraviolet light—serves as the photon entrance window. Its inner surface is coated with a high quantum-efficiency (QE) photocathode, which converts incoming photons into photoelectrons.

Below the photocathode, a microchannel plate (MCP) stack—typically configured in either Chevron or Z-stack geometry—is positioned to amplify the signal. The MCP multiplies the single photoelectrons emitted by the photocathode through secondary electron emission.

The final amplification stage consists of a bare Timepix4 CMOS ASIC, which serves as the embedded, pixelated anode. The Timepix4 is bump-bonded to a ceramic carrier board that forms the base of the vacuum tube assembly. This ceramic board serves a dual purpose: it provides mechanical sealing of the vacuum environment and offers the electrical interface between the vacuum-enclosed components and the external circuitry in air.

Signal and power transmission is handled via high-speed pins integrated into a custom Pin Grid Array (PGA), which allows the detector to be plugged into a dedicated socket on a Printed Circuit Board (PCB). The PCB connects the assembly to a custom FPGA-based data acquisition system, responsible for configuring the Timepix4 and managing data readout.

During operation, the Timepix4 can dissipate up to $5\,\watt$ of power. To ensure thermal stability and prevent overheating—which could impair detector performance—a heat sink is mounted on the ASIC and actively cooled using a chiller system.

The operating principle of the hybrid detector is as follows:
\begin{itemize}
    \item A photon enters through the input window and is absorbed by the high-QE photocathode, resulting in the emission of a photoelectron;
    \item The photoelectron is accelerated by an electric drift field toward the MCP stack, where it initiates an avalanche of secondary electrons;
    \item The amplified electron cloud is further drifted onto the bump-bonding pads of the bare Timepix4 ASIC, where it is detected by the readout electronics;
    \item The Timepix4 amplifies, discriminates, and digitizes the signal, providing precise timing and charge information.
\end{itemize}

\section{Timepix4 characterization}
Timepix4 is the latest generation application-specific integrated circuit (ASIC) of the Medipix family, developed by the Medipix4 Collaboration. It is primarily designed for single particle detection in hybrid pixel detectors~\cite{Llopart2022,sriskaran2024high}. Fabricated using 65~nm CMOS technology, Timepix4 achieves high spatial and temporal resolutions through its fine 55~\textmu m pixel pitch and a Time-to-Digital Converter (TDC) with a bin size of 195~ps. Additionally, energy measurements with a resolution of approximately 1~keV are enabled via the Time-over-Threshold ($\textrm{ToT}$) technique.

The ASIC features around 230,000 individual pixels arranged in a 448$\times$512 matrix, each equipped with both analog and digital front-end electronics. It adopts a data-driven readout architecture capable of sustaining data bandwidths up to 160~Gb/s. To handle this data throughput, FPGA-based external electronics are employed, serving not only for high-speed data acquisition but also for flexible configuration and control of the ASIC.

A comprehensive characterization of the Timepix4 ASIC—particularly with respect to timing resolution—is essential for assessing the performance of the hybrid MCP-based photon detector currently under development. Achieving the expected performance requires meticulous calibration procedures. These include threshold equalization, performed via fine tuning of the local thresholds on a per-pixel basis, and a per-pixel calibration of the ToT response as a function of deposited charge.

Timing resolution measurements were conducted using a test assembly where Timepix4 was bump-bonded to a 100~\textmu m thick n-on-p silicon sensor~\cite{akiba2025timepix4beamtelescope}. The breakdown voltage is $-180\;\volt$, and for the measurements here described the sensor has been set at $-150\;\volt$. On the sensor back-side a metallization is coated, featuring a pattern of circular openings with $300\;\micro\meter$ diameter disposed in a triangular mesh with $1\;\milli\meter$ pitch. The sensor was illuminated by a picosecond infrared laser~\cite{bib:Bolzonella_2024}. The setup involved sending simultaneous signals to multiple pixels and measuring the time difference between them to evaluate timing resolution. One pulse was sent directly to a digital pixel on Timepix4, while the second pulse triggered a pulsed diode laser (PicoQuant PDL-800 B driver with LDH-P-1060 laser head). The infrared pulses, generated with an energy of $40\,\pico\joule$, were attenuated and delivered to the sensor through an optical fiber patch cord. The laser attenuation has been varied between 15 dB and 30 dB, exploiting multiple photons detection to estimate the timing resolution for different input charges, covering a range between $1\,\kilo\textrm{e}^-$ to $200\,\kilo\textrm{e}^-$ per pixel, resulting on a total cluster charge reaching up to $\sim3200\,\kilo\textrm{e}^-$.

Timepix4 timing resolution is significantly influenced by variations in the oscillation frequency of the Voltage-Controlled Oscillator (VCO) across the pixel matrix. An automated calibration procedure was implemented to correct for these variations across all pixels. Furthermore, an additional per-pixel correction was applied to compensate for time walk effects.

\begin{figure}[htbp]
\centering
\includegraphics[width=.4\textwidth]{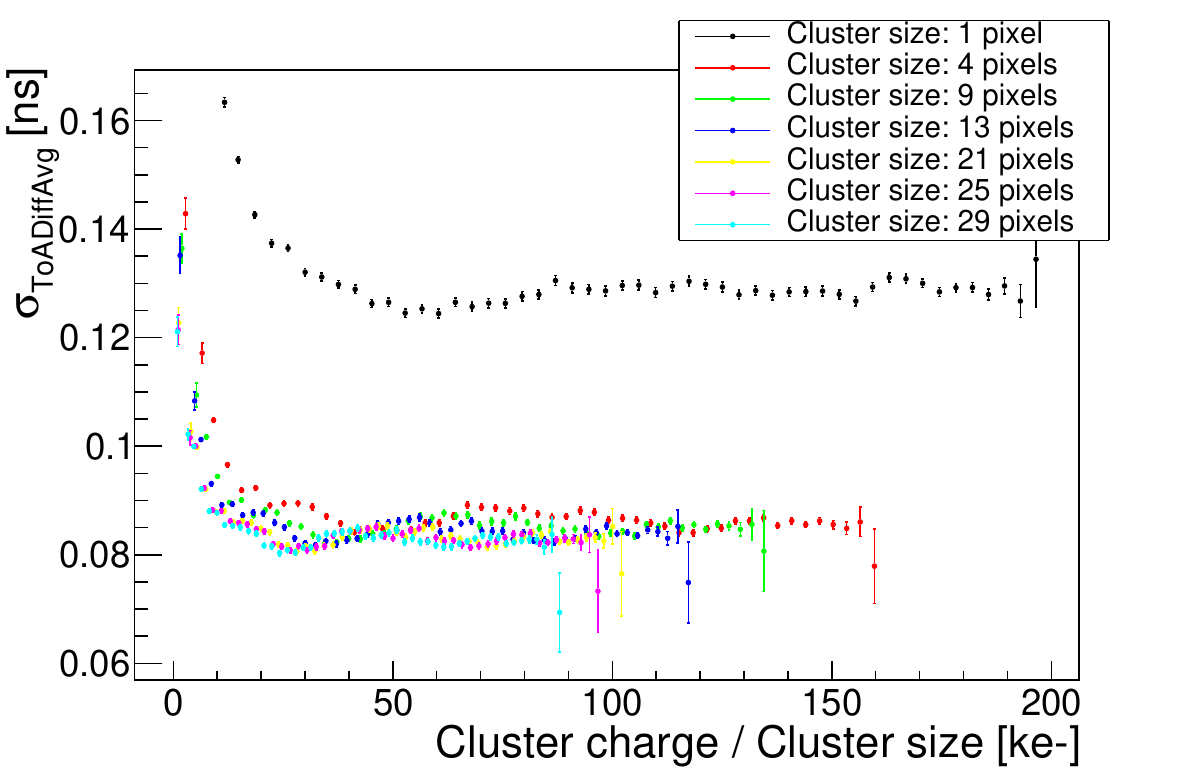}
\caption{Cluster timing resolution as a function of the cluster charge, normalized by the cluster size. Different colors correspond to different cluster sizes. The timing resolution of the reference pixel has not been subtracted~\cite{bib:Bolzonella_2024}. \label{fig:res_shell_dependence}}
\end{figure}

After these calibrations, the single pixel timing resolution reaches a plateau at $\sigma_{\textrm{pixel}} = 107\pm3\;\pico\second\;\textrm{r.m.s.}$ once the reference signal contribution, that has been estimated to be $\sigma_{ref} = 72 \pm 1\,\pico\second$ r.m.s., has been subtracted~\cite{bib:Bolzonella_2024}. When timing information from multiple pixels of the same cluster is combined, the resolution distribution plateau significantly decreases to $\sigma_{\textrm{cluster}} = 33\pm3\;\pico\second\;\textrm{r.m.s.}$, thanks to the benefits of multi-pixel sampling, as shown in Figure~\ref{fig:res_shell_dependence}. The cluster averaging analysis has been performed both combining all the pixels illuminated in each cluster (with an average size of order of almost 30 pixels), and selecting just smaller shells of pixels around the most illuminated one~\cite{bib:Bolzonella_2024}.

\section{MCP-PMT characterization}

The ceramic carrier boards were custom-designed by INFN and manufactured by Kyocera. A comprehensive series of electrical, mechanical, and thermal tests—including leak rate assessments and planarity verification—were carried out to ensure the carriers met the stringent requirements for the first MCP-PMT prototype assemblies.

In addition, two quality control setups were commissioned at CERN and at Hamamatsu Photonics Japan. These setups are used to characterize the ceramic carriers post-bonding with the Timepix4 ASIC. The characterization includes basic performance checks and ASIC-level diagnostics, ensuring reliable integration of the ceramic carriers with the pixelated readout.

During the second quarter of 2024, the first MCP-PMT prototypes produced by Hamamatsu Photonics (HPK) were delivered. These initial devices feature a photocathode with a peak Quantum Efficiency exceeding $30\%$ at a wavelength of $380\,\nano\meter$, and employ MCP stacks with channel diameters of $6\,\micro\meter$ and a pitch of $7.5\,\micro\meter$. Several prototype variants have been fabricated to enable a complete characterization campaign, differing by the number of MCP stages (2 or 3), and by the end-spoiling geometry (1D, 2D, or 3D depth). In the measurements presented in this paper the prototype with 3 MCP and 3d end-spoiling had been used.

To minimize background light and reduce output noise, each phototube is housed inside a dark box that suppresses unwanted photon hits on the photocathode. A dedicated cooling system connected to a chiller is used to bring the detector temperature down to approximately $0^\circ\,$C. The cooling infrastructure consists of two primary components: a hollow copper plate with circulating coolant is thermally coupled to the central area of the ceramic carrier's bottom surface; additionally, a copper hollow ring is thermally connected to the quartz entrance window to enhance cooling of the photocathode and improve heat dissipation from both sides of the MCP stack.

To prevent condensation due to cooling below ambient temperature, dry air is continuously flushed into the dark box. This maintains a relative humidity of approximately $5\%$, corresponding to a dew point of around $-20^\circ\,$C, effectively avoiding the formation of moisture on the cold surfaces.

\subsection{Dark Count Rate and Gain}

The dark count rate and the gain of the detector have been estimated through data-driven acquisitions conducted under dark conditions. The measurements have been performed under various biasing schemes by adjusting the voltage differences between key components (from $50\,\volt$ to $300\,\volt$ between MCP and Timepix4 and between photocathode and MCP, and from $2000\,\volt$ to $2400\,\volt$ between the two MCP faces).

A non-uniform spatial distribution of the dark count rate has been observed, and further investigations are ongoing to better understand this behavior. The hypothesis that electric field distortions could be responsible for the observed non-uniformity has been ruled out. This was verified using localized laser illumination at precise positions on the pixel matrix, which did not reveal anomalies attributable to field effects.

Instead, the variation in charge distribution and cluster sizes across the matrix suggests that gain non-uniformities are the most likely cause. This interpretation is further supported by the spatial characteristics of the observed signal. Ongoing studies, including detailed electric field simulations, are aimed at gaining deeper insight into these variations.

\begin{figure}[htbp]
\centering
\includegraphics[width=.45\textwidth]{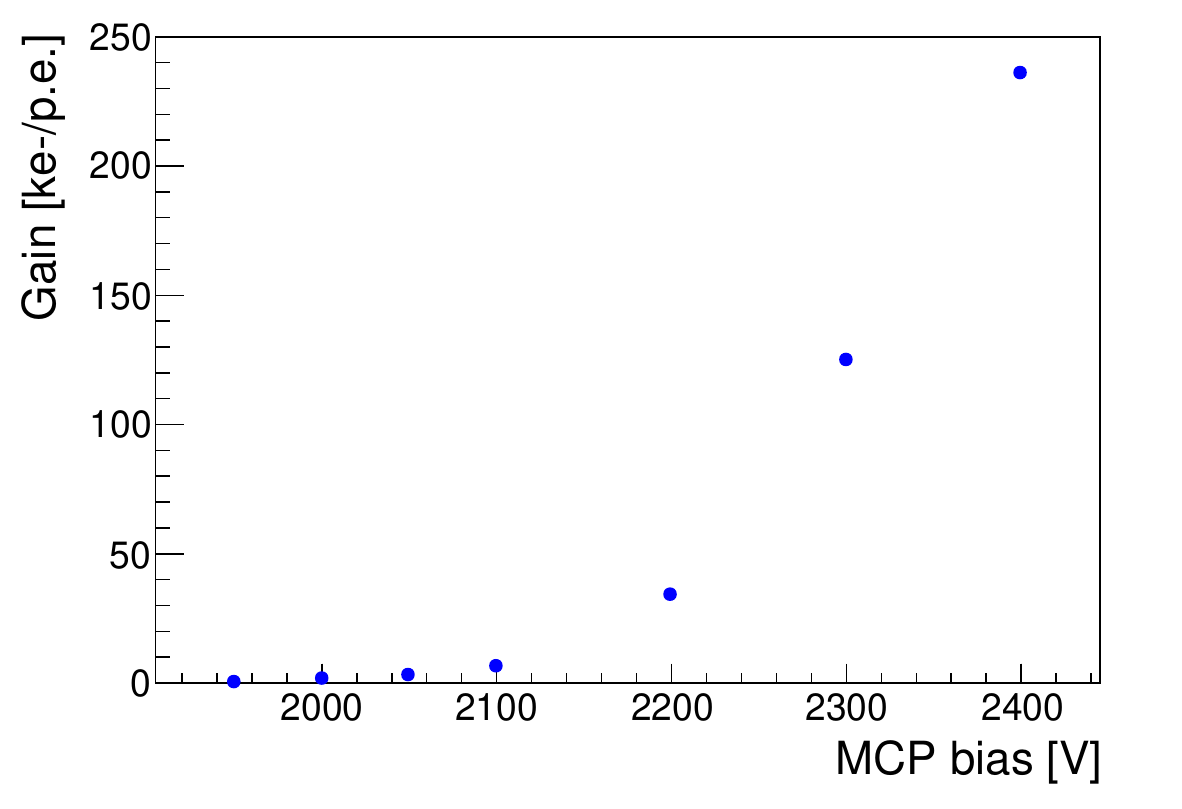}
\caption{Estimated gain as a function of the MCP bias voltage, evaluated in the central region of the Timepix4 matrix. \label{fig:central_matrix_gain}}
\end{figure}

Despite the non-uniformities, the average dark count rates are in agreement with the specifications provided by Hamamatsu, and it remains lower than $30\,\hertz\per\centi\meter^2$ when the MCP is biased at more than $2200\,\volt$. Gain values have also been estimated at various MCP bias voltages by analyzing the charge distributions in the central region of the matrix, where the spatial response is more uniform, obtaining the gain trend shown in Figure \ref{fig:central_matrix_gain}.

\subsection{Timing Resolution}

Timing resolution measurements have been carried out under conditions similar to those used with the Timepix4 bonded to a silicon sensor. To maximize performance, the MCP was operated at a gain of $240\,\kilo\textrm{e}^{-}\per\textrm{p.e.}$, biasing the MCP at $2400\,\volt$ as shown in Figure~\ref{fig:central_matrix_gain}, increasing the charge deposited across the pixels and reducing signal jitter. At the same time, electron cloud focusing was intentionally minimized to exploit the multi-sampling capabilities previously demonstrated.

The laser was highly attenuated, reaching a photon detection probability of 0.1 for a given laser pulse, to achieve single-photon operation. Under these conditions, signal clusters typically exhibited charge peaks spanning 3–4 pixels, with a surrounding low-charge halo.

Using an analysis analogous to the one described earlier, the single-pixel timing resolution reaches a plateau at approximately $95\,\pico\second$. When considering clusters composed of 4 or more pixels, the timing resolution improves to about $65\,\pico\second$.

The overall resolution is still primarily limited by the contributions from the Time-to-Digital Converter (TDC) and the reference timing signal. Furthermore, the multi-sampling benefit is less pronounced than in the silicon sensor assembly, due to the relatively small number of pixels recording significant charge within each cluster.

\section{Conclusions}

The 4DPHOTON project, funded by the ERC and hosted by INFN with support from CERN and the University of Ferrara, is developing a detector capable of measuring single photons with high spatial and timing resolution across a large active area, while maintaining low noise even at room temperature.

The detector is a "hybrid" assembly, featuring a vacuum tube housing a photocathode, a Micro Channel Plate (MCP), and a Timepix4 ASIC used as a pixelated anode with integrated readout. Photons converted into photo-electrons by the photocathode are directed to the MCP, where they are multiplied. The resulting electron clouds are then spread across the Timepix4 pixels, where the induced signal is discriminated, digitized, and read out.

Timepix4 assemblies bonded to silicon sensors have been characterized using a picosecond laser setup. The results demonstrated that the Timepix4 ASIC maintains a timing resolution below $100\,\pico\second$ and confirmed the improvement in timing resolution achieved by estimating the cluster centroid ToA compared to single-pixel ToA.

Quality test setups have been commissioned at Hamamatsu and Ferrara to characterize the MCP-PMTs. The photo-tubes are placed within a dark box and connected to an external DAQ system. Baseline tests include dark count rate (DCR) and gain measurements. Additionally, the Ferrara setup features a laser system for characterizing both timing and spatial resolution.

Several MCP-PMT prototypes with varying MCP stack numbers and end-spoiling depths have been produced. One prototype with a Z-stack configuration and 1D end-spoiling was thoroughly characterized. Results confirmed that the hybrid assembly performs as expected, with gains and dark count rates in line with the design. Timing resolutions below $100\,\pico\second$ were achieved, and further optimization of the Timepix4 analog front-end is expected to improve resolution.

The characterization process will continue on other prototypes to compare their performance and identify the optimal configuration for production.

\section*{Acknowledgments}
This work was carried out in the context of the Medipix4 Collaboration based at CERN, and in the framework of the MEDIPIX4 project funded by INFN CSN5.
This project has received funding from the European Research Council (ERC) under the European Union's Horizon 2020 research and innovation programme (Grant agreement No. 819627, 4DPHOTON project).
We acknowledge the VELO team of the LHCb collaboration for lending us the Timepix4 assembly bonded to the n-on-p silicon sensor, which allowed us to carry out the timing resolution measurements described in this paper.
  \bibliographystyle{elsarticle-num-names} 
  \bibliography{biblio.bib}
\end{document}